# Industrial Applications of Neutrinos

**Giovanna Takano Natti**[a,1], **Érica Regina Takano Natti**[b,2], **Paulo Laerte Natti**[c,3,*]

[a] College of Science and Engineering, University of Minnesota
Twin Cities, Minneapolis, MN 55455.
[b] Engineering Department, Pontifícia Universidade Católica do Paraná
Rua Jóquei Clube, 458, Londrina, Paraná, Brasil, 86067-000
[c] Mathematics Department, Universidade Estadual de Londrina,
Campus Universitário, Londrina, Paraná, Brasil, 86051-990

[1] E-mail: takan009@umn.edu
[2] E-mail: erica.natti@pucpr.br
[3] E-mail: plnatti@uel.br
[*] Corresponding author

**ABSTRACT**. We present a review of the current and future industrial applications of neutrinos. We address the industrial applications of neutrinos in geological and geochemical studies of the Earth's interior, in monitoring earthquakes, in terrestrial communications, in applications for submarines, in monitoring nuclear power plants and fusion reactors, in the management of fissile materials used in nuclear plants, in tracking nuclear tests, among other applications. We also address future possibilities for industrial applications of neutrinos, especially concerning communications in the solar system and geotomography of solar system bodies.

**Keywords:** Neutrino detectors. Neutrino communications. Geotomography. Nuclear power plant safety. Earthquake monitoring.

## 1. Introduction

Among the particles of the Standard Model of Elementary Particle Physics [1-4], neutrinos are certainly the most enigmatic elementary particles. Due to their possibly very small mass and absence of electrical charge, neutrinos interact very weakly with other elementary particles and fields. On the other hand, due to these unique characteristics, neutrinos can be used to obtain information in circumstances where other particles or fields are strongly absorbed, such as, for example, electromagnetic waves in ionized media.

We explore the possible industrial applications for neutrinos. We emphasize that the objective is not to discuss the so-called academic applications, but rather the initial attempts at industrial applications for neutrinos.

First, we present their current industrial applications, or that have already been experimentally tested, among them the monitoring of nuclear power plants and nuclear weapons, geological and geochemical studies of the Earth's interior, terrestrial communications, development of small-sized neutrino detectors, and development of neutrino beam generators. We also present and discuss possible future industrial applications of neutrinos, such as the improvement of terrestrial and interplanetary communication by neutrinos, applications for submarines, monitoring of spent nuclear fuel, monitoring of nuclear tests, monitoring of fusion reactors, geotomography of solar system bodies, and earthquake forecast.

## 2 – Neutrinos

The main historical facts and discoveries about neutrinos (electron neutrino, muon neutrino, and tau neutrino) can be found in [5]. The most recent measurements of neutrino properties are presented in [1]. This section describes the methods for neutrino detection and the main neutrino observatories.

### 2.1 – Detection of Neutrinos

Three known processes make neutrino detection possible. These processes are described below.

2.1.1 - $W^-$ Charged Current Reaction

In this process, when an electron neutrino approaches the nucleus of a deuterium $D$, the electron neutrino exchanges a charged particle of the weak force, a $W^-$ boson. In this interaction, the deuterium neutron becomes a proton, and the electron neutrino becomes an electron, that is,

$$\nu_e + D \rightarrow p + p + e^- \quad \text{or} \quad \nu_e + n \rightarrow p + e^-.$$

Note that the energy of the incident neutrino is very high and, consequently, the speed of the emitted electron is close to the speed of light in vacuum, approximately $c = 3 \times 10^5$ m/s. However, this electron is emitted in a dielectric medium, in a tank of heavy water $(D_2O)$, where the speed of light is around $2.2 \times 10^5$ m/s. When a charged particle crosses a dielectric medium with a speed greater than the speed of light in this medium, it causes shock waves (like acoustic shock waves) that emit blue radiation, the so-called Cherenkov radiation [6]. From the patterns of emitted Cherenkov radiation, the energy and angular distribution of neutrino can be determined. The $W^-$ charged current reaction is equally sensitive for the three flavors of neutrinos.

2.1.2 - $Z^0$ Neutral Current Reaction

In this process, an electron neutrino approaching the deuterium nucleus exchanges a neutral particle of the weak force, a boson $Z^0$. In this interaction, due to the scattering of the electron neutrino by deuterium, the fission of the deuterium nucleus into a proton $p$ and a neutron $n$ occurs, that is,

$$\nu_e + D \rightarrow n + p + \nu_e \ .$$

Furthermore, the emitted neutron is captured by another nucleus, for example, by the nucleus of a chlorine atom, added to heavy water in the form of NaCl. The Cl nucleus, upon absorbing the neutron, emits gamma rays that will scatter electrons, which will produce Cherenkov light. The $Z^0$ neutral current reaction is equally sensitive to the three types of neutrinos.

2.1.3 - $\gamma$ Electron Scattering Reaction

The so-called $\gamma$ electron scattering reaction by neutrinos does not only happen in heavy water ($D_2O$) nuclei, it also occurs in ordinary water ($H_2O$) detectors. In this process, electrons bound to hydrogen and deuterium atoms are scattered with very high energies by the incident electron neutrino, as shown in the scattering reaction below,

$$\nu_e + e^- \rightarrow \nu_e + e^- \ .$$

As in the processes above, scattered electrons emit Cherenkov Light. Although the electron scattering reaction is sensitive to all flavors of the neutrino, the one with the electron neutrino occurs six times as often.

## 2.2 – Neutrino Observatories

The three largest neutrino observatories are the Super-Kamiokande Observatory, the Sudbury Neutrino Observatory, and the IceCube Neutrino Observatory.

The Super-Kamiokande Observatory (SKO) [7] is located 1 km deep in a mine in Mozumi, Japan. In operation since April 1996, it is a cylindrical tank measuring 39 m in diameter and 42 m in height that contains 50,000 tons of ultrapure water. Inside there are 13,000 photomultiplier tubes that detect Cherenkov radiation. The Super-Kamiokande Collaboration announced the first evidence of neutrino oscillation in 1998, supporting the theory that the neutrino has a non-zero mass. In recent decades, experiments at SKO have been designed to detect solar and atmospheric neutrinos and search for neutrinos from supernovae in the Milky Way. The Hyper-Kamiokande Observatory (HKO), the successor to the Super-Kamiokande, is being planned.

The Sudbury Neutrino Observatory (SNO) [8] was a neutrino observatory located 2,1 km deep in a mine in Sudbury, Ontario, Canada. The detector was a sphere with a radius of 6 meters, containing 1,000 tons of heavy water and 9,600 photomultiplier tubes (PMTs). It was in operation from May 1999 until November 2006. It was designed to detect the oscillations of solar neutrinos. In June 2001, the SNO presented the first clear evidence that solar neutrinos oscillated as they traveled to Earth. The SNO+ is the successor observatory to the SNO and, among other experiences, is designed to detect geoneutrinos, neutrinos from nuclear reactors, and neutrinos from supernovae.

The IceCube Neutrino Observatory (INO) [9] was completed in December 2010 at the Amundsen-Scott South Pole Station in Antarctica. The INO is the largest neutrino telescope in the world. Its thousands of sensors are located under the ice, deployed at depths between 1,450 and 2,450 meters, spread over a cubic kilometer. IceCube is designed to detect high-energy astrophysical neutrinos, in the range of $10^7$ eV to about $10^{21}$ eV. In July 2018, INO announced that they had tracked an extremely high-energy neutrino to its point of origin in the blazar TXS 0506 +056, located 3.7 billion light-years away. This is the first time a neutrino detector has identified an extrasolar source for an observed neutrino.

Like the INO project, in the Mediterranean Sea, there are several submerged neutrino observatories, among which we mention ANTARES (telescope working since May 2008 on the coast of France), NEMO (coast of Italy), and NESTOR (coast of Greece). These last three projects were incorporated by the Cubic Kilometre Neutrino Telescope (KM3NeT), a European project to build a large underwater neutrino observatory.

Some other large operational neutrino observatories are listed below: Baksan Neutrino Observatory (BNO) in a mine in Baksan, Russia; Baikal Gigaton Volume Detector (Baikal-GVD) in Lake Baikal, Russia; China Jinping Underground Laboratory (CJPL) in Sichuan, China; Daya Bay Reactor Neutrino Experiment in Daya Bay, China; FermiLab in Illinois, US; Laboratori Nazionali del Gran Sasso (LNGS) in the Gran Sasso mountain, Italy; Laboratorio Subterráneo de Canfranc (LSC) in the Aragon Pyrenees, Spain; Modane Underground Laboratory (LSM) in Moldane, France; Pierre Auger Observatory in Mendoza, Argentine; and Telescope Array in Utah, US.

There are also several experiments with smaller neutrino detectors, for instance, placed near nuclear power plants or in atmospheric balloons.

## 3 – Industrial Application of Neutrinos

We review current and future industrial applications of neutrinos. First, current industrial applications that are already being tested experimentally are described. Then, future industrial applications that have not yet been implemented are discussed.

### 3.1 - Current industrial neutrino applications

3.1.1 - Monitoring of Nuclear Power Plants and Weapons

In the chamber of a nuclear reactor, controlled nuclear fission generates energy. These cameras are shielded, so no radiation should escape from them, except neutrinos. The observation of neutrinos that emerge from nuclear power plants provides a diagnosis of the reactor, information about the plant's safety, efficiency, and the amount of radioactive elements produced, including plutonium, which can be used for the development of nuclear weapons. Therefore, by observing the neutrino flux from a nuclear power plant, it is possible to verify whether the current International Atomic Energy Agency (IAEA) safeguards are being satisfied. It is also possible to discover or monitor non-cooperative nuclear reactors [10-14].

In this context, since the 1980s, the construction of compact neutrino detectors has been improved. By compact detectors, we mean detectors that can be transported. There are currently several compact neutrino detectors in operation, located near nuclear power plants, which aim to monitor their nuclear reactors. In general, these detectors are scintillation type, using liquid or solid materials doped with Lithium or Gadolinium. We cite some of these detectors: CHANDLER [15], CONUS [16], DANSS [17], Neutrino-Angra [18], Nucifer [19], NuLat [20], PROSPECT [21], SANDD [22], SoLID [23], VIDDAR [24], WATCHMAN [25].

3.1.2 - Geotomography of the Earth

The oscillation between neutrino flavors is affected by the type of matter in which the neutrino propagates. In this context, applying the neutrino oscillation phenomenon to analyze the Earth's interior would allow the detection of cavities and faults, underground water reservoirs, and oil, gas, and mineral reservoirs, among other structures [26]. Very little is known about the Earth's interior. The idea of studying the Earth's interior using neutrino tomography has been around for decades. Experiments are currently underway with this aim [26-29]. The industrial use of the neutrino geotomography methodology will allow great scientific, technological, and economic advances.

Neutrino absorption tomography uses the fact that the neutrino cross section increases with energy. Note that neutrinos with energy around 40 TeV have absorption lengths comparable to the diameter of the Earth. Consequently, the absorption of neutrinos along their straight paths through the Earth can be used to study the density profile of Earth like the X-ray tomography technique [26,28,29]. In contrast, neutrino oscillation tomography of the Earth can be performed by neutrinos with energy around GeVs. Thus, neutrinos with energies in the GeV range can be used to provide measurements of the composition of the Earth's interior [26,27,29,30].

### 3.1.3 - Terrestrial Communications by Neutrinos

Since data transmission via neutrinos does not require retransmission by satellites and does not suffer power losses (interference, attenuation and dispersion), neutrino communication has been proposed for several scenarios, for example, where radio waves or optical signals are not viable, as they are strongly damped or do not propagate [31]. Among the most intuitive possibilities for neutrino communication, we highlight point-to-point terrestrial communications, submarine communications, and spatial communications. While the last two mentioned possibilities for neutrino communication are still theoretical ideas, the point-to-point terrestrial communication by neutrinos has already been realized.

In 2012, researchers at Fermilab encoded the message "neutrino" into a beam of neutrinos and sent this data to a detector 1035 meters away, including 240 meters through the Earth [32]. The neutrinos were emitted by the NuMI particle accelerator (Neutrinos at Main Injector) and captured by the MINERvA detector (Main Injector Neutrino ExpeRiment for v-A) at FermiLab [33]. The "neutrino" message was sent using ASCII code via a series of neutrino pulses (bits). It took two hours to transmit the bits and about 6 minutes to decode them. Data transmission via neutrinos was successful, although the connection maintained was very slow, around 0.1 bits per second. This experiment demonstrated the possibility of neutrino data transfer via advanced technology.

### 3.1.4 - Development of Small-Sized Neutrino Detectors

The development of smaller and more efficient neutrino detectors is essential for both nuclear power plant monitoring and terrestrial neutrino communication systems, among other current and future applications.

In the case of monitoring nuclear power plants, neutrino detectors of a size compatible with the task already exist [15-25]. In the case of point-to-point terrestrial communications by neutrinos, the NuMI particle accelerator and the MINERvA detector, which were used in the first neutrino communication experiment [32], have dimensions that are not yet compatible with industrial terrestrial communications applications. On the other hand, they would be compatible with a communication application between the Earth and the far side of the Moon, where NASA's Lunar Crater Radio Telescope (LCRT) is planned to be built.

In this context, the smallest neutrino detector ever built is the COHERENT detector [34] at Oak Ridge National Laboratory (ORNL), Tennessee. The COHERENT detector is a 14.6 kg cesium iodide scintillator crystal, doped with sodium to enhance the observation of Cherenkov radiation. A single Hamamatsu R877-100 photomultiplier detects scintillation light. These components and electronics are protected by a plastic scintillator veto to avoid activity induced by other neutrino sources.

### 3.1.5 - Development of Neutrino Beam Generators

Analogously to the need to develop smaller neutrino detectors, there is also a demand for the development of smaller neutrino beam generators. The technology for generating neutrino beams in particle accelerators has existed for decades. Since the 1960s, the Van der Meer horn has been used to focus beams of pions generated by the collision of high-energy proton beams. Then, when the pions decay into muons and neutrinos, a focused beam of neutrino is obtained. Currently, a new methodology for producing neutrino beams is being tested, one that will produce neutrino beams from the decay of muons confined within a storage ring. This new technology is being tested in the STORM [35] and ENUBET [36] experiments, both at CERN, providing monitored neutrino beams with known flux and flavor composition to within 1% accuracy.

## 3.2 - Future Industrial Applications of Neutrinos

For the possibilities presented in this section to be applicable, significant development of neutrino detectors and neutrino beam generators is required. Despite the technological limits, the proposals presented below are feasible. We believe that such future industrial applications of neutrinos will only be a matter of time.

3.2.1 - Improvement of Fast Communication by Neutrinos

The use of neutrinos for point-to-point terrestrial communications through the Earth's interior is a very interesting scenario. In this case, the fast transmission of point-to-point communications, for example, from Brazil to Japan, without retransmission by satellites, without climatic inconveniences, and without power loss would be a great advance for terrestrial communications.

Another possibility would be the use of neutrinos for communication between a lunar base on the far side of the Moon and the Earth. In the case of communication via electromagnetic waves, a system of satellites around the Moon would be necessary to relay this communication. In the case of a neutrino communication system, the communication would be directly through the Moon. In this context, we highlight the project to build the Lunar Crater Radio Telescope (LCRT) on the far side of the Moon.

A more futuristic application of neutrino communication is one in which such particles would be used for interstellar communication [37-38]. In such situations, communication by electromagnetic waves is attenuated and even blocked due to the presence of matter or strong electromagnetic fields. In the context of interstellar communication, neutrino communication would be much more efficient.

Finally, there is the possibility that advanced civilizations are using neutrino-encoded signals to communicate over interstellar distances. Note that the large neutrino observatories have projects in which they look for structured neutrino signals, which would be one way to provide evidence of extraterrestrial intelligent life [39-40].

### 3.2.2 - Applications for Submarines

Communication with submarines, or between submarines, is complex. Communication via electromagnetic waves is inadequate since these waves are strongly absorbed in ionized media. In particular, communication via radio waves of frequencies on the order of a few kHz is possible, but this technique requires a nearby antenna floating on the surface of the sea, imposes major limitations on the submarine's ability to move. On the other hand, this communication could occur at any depth, or distance between the submarine and the emitting source, if the communication were via neutrinos [41]. In this case, the submarine would need to have, in addition to the neutrino detector, a neutrino source. Nuclear submarines have nuclear reactors and are candidates for the development of this technology.

There is also a need to improve current navigation technologies for submarines. Studies are currently underway to develop Neutrino PNT (Position, Navigation and Timing) systems for underwater navigation [42]. These projects involve a global network of neutrino sources, which must be detected by submarines, together with a high-performance synchronized clock system. Improving directional neutrino detection using small detectors is the main challenge for implementing this navigation technology.

In this context, the recent development of detectors sensitive to Coherent Elastic Neutrino-Nucleus Scattering (CEvNS) has allowed the construction of smaller and more efficient detectors [34,43,44]. CEvNS occurs when a neutrino coherently interacts with the entire nucleus, which generates a small recoil of the nucleus. For neutrinos with energy below 50 MeV, a very large cross section is observed compared to non-coherent scattering at the same energy. Furthermore, other cases of interest for neutrino PNT systems are subsurface PNT, back-up of Global Navigation Satellite System (GNSS) in denied environments, and spacecraft positioning or tracking [42].

Another possible industrial application for neutrinos would be to monitor fissile material on board nuclear-powered submarines [45]. Recent agreements to transfer nuclear submarines to non-nuclear countries have raised questions about how to safeguard naval reactor fuel. Current safeguarding technologies and practices are difficult to implement in this case. On the other hand, from submarine patrols with neutrino detectors, it would be possible to verify that the nuclear core of the submarine under inspection has not been diverted or if its enrichment level has changed [45].

### 3.2.3 - Monitoring of Spent Nuclear Fuel

Mainly due to nuclear power plants, radioactive waste storage has increased rapidly over the last 70 years. Spent nuclear fuel (SNF) is stored in casks. Currently, there is no technique to monitor the contents of these storage casks. On the other hand, several studies propose the use of detectors sensitive to Coherent Elastic Neutrino-Nucleus Scattering (CEvNS) to monitor the contents of these SNF storage casks or to locate underground "hot spots" in areas contaminated by radioactive material storage [13-14,46–47]. Such measurements would make it possible to detect leaks or containment problems in these storage casks, ensuring the long-term safety of these radioactive wastes.

3.2.4 – Monitoring of Nuclear Tests

Improvements in the sensitivity of new neutrino detectors and the monitoring of earthquakes around the world by the International Seismic Network (ISN) will provide greater sensitivity in the international monitoring of undeclared nuclear weapons tests. In this context, the observation of a neutrino pulse correlated with a coincident seismic signal could provide confirmation of a nuclear explosion. Note that both underground and underwater nuclear explosions would be detected by this detection system [48].

Furthermore, with the improvement of neutrino detectors and the ISN, it is possible to think about quantifying the yield of a nuclear explosion, which would make it possible to distinguish nuclear fission weapons tests from nuclear fusion weapons tests [13-14,49].

3.2.5 - Monitoring of Nuclear Fusion Reactors

Neutrinos are also generated in nuclear fusion reactions, such as those occurring in the Sun and tokamaks. The ability to detect and measure neutrinos produced in controlled nuclear fusion processes would provide information about the efficiency and performance of these reactors. This data could be used to tune and optimize nuclear fusion processes for power generation, thereby improving the energy efficiency and safety of these fusion reactors.

Currently, nuclear fusion reactors for power generation are still in the research and development stages, as they have not yet reached the stage of commercial viability. Several fusion reactor projects are in advanced stage of development. The JET (Joint European Torus), located in Culham, United Kingdom, began operation in 1983 and is currently the largest tokamak in the world. The ITER (International Thermonuclear Experimental Reactor), located in Cadarache, France, is expected to start operating around 2025. The SPARC (Soon as Possible Affordable Robust Compact) fusion reactor, located in the United States, has the ambition of being the first commercial reactor to produce more energy than it consumes. We also mention the EAST (Experimental Advanced Superconducting Tokamak) and the HL-2M fusion reactors in China; the Wendelstein 7-X in Germany, the JT-60SA in Japan, and the KSTAR (Korea Superconducting Tokamak Advanced Research) in South Korea.

### 3.2.6 - Geotomography of Solar System Bodies

Many studies of the Earth's interior using neutrino geotomography have already been carried out [26-30]. The idea is to apply the oscillation between neutrino flavors to find internal structures and faults; oil, gas, and mineral reservoirs; water reservoirs; etc.

Similarly, in the more distant future, one could consider applying neutrino geotomography to other bodies in the solar system, such as other planets, moons, asteroids, and comets [26,50]. Neutrino geotomography of bodies in the solar system would enable major scientific, technological, and economic advances.

Finally, geoneutrinos may answer a long-standing question about the Earth's liquid core and the heat that constantly flows through the Earth's surface. How is this energy generated? What proportion of this energy is generated by radioactive decays of the $^{232}$Th, $^{238}$U, $^{235}$U, and $^{40}$K? Measurements of geoneutrino fluxes provide information about the amounts and proportions of these radioactive elements in the Earth's interior [51].

### 3.2.7 - Earthquake Forecast

We currently do not have an efficient way to predict earthquakes. Earthquakes are caused by rapid slippage along fault zones, where enormous deformation and energy release occur, caused by shear and tension. Therefore, studying the propagation of neutrino beams through the Earth's inner layers would make it possible to detect subtle changes in the density, geometry, and dynamics of fault zones. Measurements over time of these parameters (density, geometry, and dynamic deformations) on a fault could provide a methodology for predicting potential earthquakes [52].

## 4 – Discussion

We identify three main possibilities for industrial applications of neutrinos: monitoring, tomography, and communication.

Regarding practical applications for neutrinos, the monitoring and safety of nuclear reactors are the most studied topics and those that receive the largest investments. In this context, many detectors have been developed based on different technologies [15-25,34,43,44], we highlighted the inverse beta decay and Coherent Elastic Neutrino-Nucleus Scattering (CEvNS). Related to these neutrino physics applications, other possible and important neutrino monitoring applications were thought of, such as the monitoring of fissile material aboard nuclear-powered submarines, the monitoring of spent nuclear fuel depots, the monitoring of undeclared nuclear materials or facilities, the monitoring of undeclared nuclear tests or weapons, the monitoring of nuclear fusion reactors, and the monitoring of potential earthquakes.

Another important possibility for the applications of Neutrino Physics are the studies of geotomography of the Earth and the future applications of neutrino geotomography to other bodies in the solar system. Geotomography of the Earth and other bodies in the solar system will allow great scientific, technological, and economic advances.

The third potential major industrial application of neutrinos is fast point-to-point neutrino communication. Terrestrial communications and communications with submarines, or between submarines, would be greatly facilitated by this new technology. This communication technology is also promising and will become increasingly necessary as the frontiers of our civilization expand. Communication within our Solar System would already benefit greatly from the use of this technology. On the other hand, the use of neutrino communication will be essential for future spacecraft sent outside the Solar System, as in the case of the Voyager 1 and 2 spacecraft.

Finally, we emphasize that for the possible industrial applications of neutrinos presented in this review to be viable, significant technological developments in the sensitivity of these neutrino detectors and the development of smaller neutrino beam generators will be necessary.

## Statements and Declarations

Funding Declaration: The authors declare no funding.

Competing Interest Declaration: The authors declare no competing interest.

Author Contribution Declaration: All authors wrote the manuscript text and revised it.

## Ethics Declaration

The work presented is original and has not been published anywhere else, in any form or language.